\documentclass{eptcs}
\providecommand{\event}{TTC 2011}

\usepackage{oz2}
\usepackage{alltt}
\usepackage{graphicx}

\title{Solving the TTC 2011 Model Migration Case with UML-RSDS}
\author{K. Lano, S. Kolahdouz-Rahimi\\
{Dept. of Informatics, King's College London, Strand, London, UK\thanks{Research supported by the HoRTMoDA EPSRC project}}
\email{kevin.lano@kcl.ac.uk}}

\def\titlerunning{Model Migration Case in UML-RSDS}
\def\authorrunning{K. Lano, S. Kolahdouz-Rahimi}

\begin{document}
\maketitle

\begin{abstract}
In this paper we apply the UML-RSDS notation
and tools to the GMF model migration 
case study and explain how to use the UML-RSDS tools.
\end{abstract}

\section{Model transformation specification in UML-RSDS}

UML-RSDS is a model-driven development method
with an associated toolset. It was originally designed
as a general-purpose method for synthesising 
verified executable systems from high-level 
specifications \cite{Lano08cdd}, and has been adapted
for the synthesis of transformation implementations
from specifications \cite{Lano11icmt}. Modelling is
carried out using UML 2: class diagram models, use
cases, state
machines, activities, object models and interactions.

In UML-RSDS the specification of a 
transformation is written in
first-order logic and OCL, defining the preconditions
(assumptions $Asm$) of the use case representing the
transformation, and the postconditions $Cons$ of the
use case. Transformations may be composed using
chaining and the $includes$ and $extends$ composition
mechanisms of UML use cases.

\section{GMF model migration}

This case study \cite{modelmigrationcase}
is a re-expression transformation which
involves a complex restructuring of the data of a 
model: actual figures are replaced by references to
figures, and references from a figure to subfigures
are recorded by explicit objects.

\begin{figure}[htbp]
\centering
\includegraphics[width=4.1in]{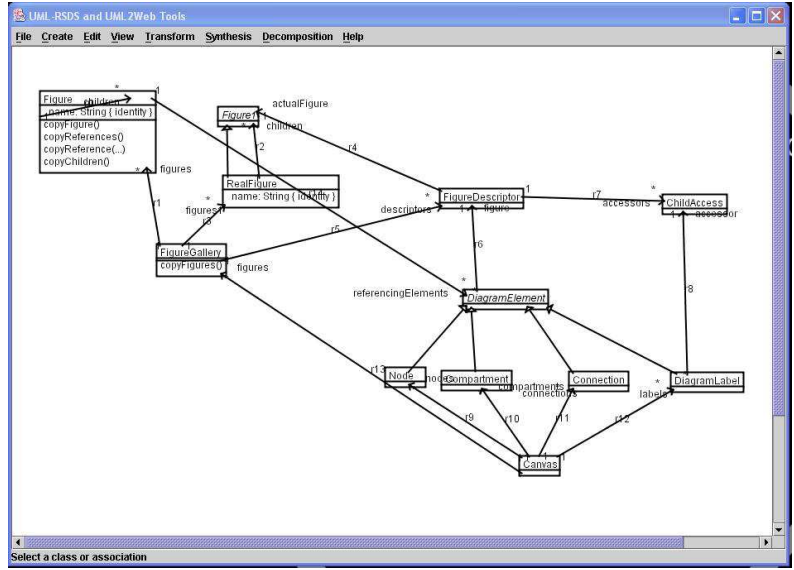}
\caption{GMF metamodels in UML-RSDS}
\label{gmfmm}
\end{figure}

Figure \ref{gmfmm} shows the unified metamodels of
the source (GMF version 1.0) and target (GMF version 2.1)
languages. Since most of the data of a model may 
remain unchanged by the transformation, we specify
the transformation as an update-in-place mapping.
$Figure1$ is the target metamodel
version of the $Figure$ class,
$figures1$ is the target version of the gallery figure list
association end. 

Class diagrams can be created using the visual
class diagram editor of the UML-RSDS tool
(executed by invoking {\tt java UmlTool}).

We assume in $Asm$ that the input model is a syntactically
correct version 1.0 model and that the new entities have
no instances:
\[ Figure1 = \{\} \\
FigureDescriptor = \{\}\\
ChildAccess = \{\} \]

For simplicity of specification, we decompose the
transformation into a first transformation which
creates the new data from the old, without deleting
any data, and a second transformation which removes
the version 1.0 data which is not in version 2.1.
This is an example of the {\em construction and cleanup}
design pattern \cite{Lano11icsea}.

For clarity, we use conventional mathematical
notation here, the specification must
however be written in the ASCII syntax for OCL
when entered into the toolset (Appendix \ref{bnf}).

The first transformation is specified by the following 
$Cons$ constraints:
\[ (C1): ~ \\
\forall f : Figure \cdot \exists_1 rf : RealFigure \cdot rf.name = f.name ~~and\\
\t2 ~~~\exists_1 fd : FigureDescriptor \cdot fd.actualFigure = rf \]
For each source model figure, there is a unique
target model
real figure, with a figure descriptor.

\[ (C2):\\
\forall f : Figure \cdot RealFigure[f.name].children = RealFigure[f.children.name] \]
For each source model figure, the target model
real figure has children the corresponding children
(real figures). The notation $E[x]$ denotes the instance of
$E$ identified by the key value $x$, if this is a single
value, or the set of $E$ instances identified by $x$, if
$x$ is a set.

\[ (C3):\\
\forall fg : FigureGallery ~\cdot \\
\t2 fg.figures1 = RealFigure[fg.figures.name] ~~and\\
\t2 fg.descriptors = FigureDescriptor{\fun}select(actualFigure : fg.figures1) \]
For each figure gallery, its figures ($figures1$)
in the target model are the 
real figures corresponding to the source model
figures of the gallery, its descriptors are the 
descriptors of these figures. Although in this 
constraint $figures1$ is both written and read, the
update only affects the local data of one
$FigureGallery$ object $fg$, and no
other object is modified, so no other application of
the rule is affected.

\[ (C4):\\
\forall f : Figure; fd : FigureDescriptor; d : f.referencingElements ~\cdot \\
\t1 fd.actualFigure = RealFigure[f.name] ~~ implies\\
\t2    d.figure = fd ~~and \\
\t2    (d : DiagramLabel ~~implies\\
\t3       \exists ca : ChildAccess ~\cdot~ d.accessor = ca ~~and~~ ca : fd.accessors) \]
The figure descriptor of a diagram element in the target
model is that corresponding to the figure which
contained the element in the source model. If the
diagram element is a label of a nested figure, then an
explicit child access object is defined to record the 
access (\cite{modelmigrationcase}, page 3).


Each of the $Cons$ constraints 
can be implemented by simple
iterations \cite{Lano11icsea}. This implementation is
carried out automatically by the UML-RSDS toolkit: a
design level description as a UML activity is derived
for each use case. In addition, executable Java code is
also generated. The implementation is structured as a
sequence of phases, one for each constraint.
The phase for
$C1$ must precede the phases for the other
three constraints, but they can be executed in any order,
so the transformation can be decomposed into several
separate use cases if required. Only $C4$ uses the
$DiagramElement$ class and its subclasses, so an
input model could be divided into two parts, with
the instances of
classes $Figure$, $FigureGallery$ required for $C1$ to
$C3$, and instances of the other classes required for
$C4$.

$C1$ and $C2$ are implemented by
iterations over $Figure$
of operations $copyFigure$ and $copyChildren$,
respectively. $C3$ is implemented by an iteration
of an operation $copyFigures$ over $FigureGallery$.
$C4$ is implemented by an iteration of an
operation $createReferences$ on $Figure$.

The BNF syntax of the OCL subset used in UML-RSDS
is defined in Appendix \ref{bnf}. Metamodels are
stored in text files in the $output$ subdirectory, but
should not be edited directly, only via the graphical
editor of UML-RSDS.

The second transformation
removes all instances of
$Figure$ and all elements and links specific
to the source metamodel. It is an update-in-place
transformation, with $Cons$ specification
\[ Figure{\mbox{@}pre}.referencingElements = \{\}\\
FigureGallery.figures = \{\}\\
Figure{\fun}isDeleted()
\]
This can be coded as the postcondition of an operation
$cleanModel$ of $Canvas$.

The two transformations are composed by executing 
one after the other, using an intermediate file to
hold the target model of the first transformation,
which serves as the source model of the second.

\section{Conclusion}

We have shown that UML-RSDS can specify the GMF
case study transformation in a direct and
concise manner, both
as high-level specifications and as explicit designs.
UML-RSDS has the advantage of using standard UML
and OCL notations to specify transformations, reducing
the cost of learning a special-purpose transformation
language. Our method has the advantage of making
explicit all assumptions on models 
and providing global specifications ($Cons$ and 
$Asm$) of transformations, independent of specific
rules.

One deficiency is a lack of graphical specification 
for transformation rules, ie, by diagrams at the
abstract or concrete syntax level. We intend to
support such specification as a supplement to the
formal specifications of rules.

\appendix

\section{Transforming specific models}

Source and target metamodels are defined using the
visual class diagram editor of UML-RSDS. Metamodels
cannot contain multiple inheritance, and all
non-leaf classes must be abstract. Metamodels can be
saved to a file by the $Save$ $data$ command, and
loaded by $Load$ $data$.

Source models can be
defined in text files, which are then read by
the executable implementation $Controller.class$
of the transformation, 
in a textual form. An example is shown
below for GMF.

UML-RSDS can be executed by the command 
{\tt java UmlTool}. The directory {\tt output} is used to
store metamodels, input and output
models, and the generated Java code. The command
$Load~data$ loads a metamodel from a file
(eg, $gmfmm3.txt$ for the migration metamodel).
The command $Synthesis~Java$ generates the Java
executable of a transformation, this
generated executable is the 
{\tt Controller.java} file in the $output$ directory.
This can be compiled and used independently of the
toolset. It is compatible with Java SDK version 1.4.1 and
later versions, the only specialised Java package used
is Java reflection, to load models.

An example source model ($gmf1.txt$) is as follows:
\begin{small}
\begin{verbatim}
c : Canvas
c1 : Compartment
c2 : Compartment
c1 : c.compartments
c2 : c.compartments
n1 : Node
n2 : Node
n1 : c.nodes
n2 : c.nodes
l : DiagramLabel
l : c.labels
fg : FigureGallery
fg : c.figures
f1 : Figure
f1.name = "f1"
f2 : Figure
f2.name = "f2"
f2 : f1.children
f1 : fg.figures
l : f1.referencingElements
n1 : f1.referencingElements
c1 : f1.referencingElements
n2 : f2.referencingElements
c2 : f2.referencingElements
\end{verbatim}
\end{small}
The new model generated from this is:
\begin{small}
\begin{verbatim}
c : Canvas
c1 : Compartment
c2 : Compartment
c1 : c.compartments
c2 : c.compartments
n1 : Node
n2 : Node
n1 : c.nodes
n2 : c.nodes
l : DiagramLabel
l : c.labels
fg : FigureGallery
fg : c.figures
rf1 : RealFigure
rf1.name = "f1"
rf2 : RealFigure
rf2.name = "f2"
rf2 : rf1.children
fd1 : FigureDescriptor
fd1.actualFigure = rf1
fd2 : FigureDescriptor
fd2.actualFigure = rf2
rf1 : fg.figures1
fd1 : fg.descriptors
l.figure = fd1
n1.figure = fd1
c1.figure = fd1
n2.figure = fd2
c2.figure = fd2
ca : ChildAccess
l.accessor = ca
ca : fd1.accessors
\end{verbatim}
\end{small}

\section{Expression syntax of UML-RSDS}
\label{bnf}

UML-RSDS uses both classical set theory expressions and
OCL. It only uses sets and sequences, and not bags or
ordered sets, unlike OCL. Symmetric binary operators
such as $\cup$ and $\cap$ are written in the classical
style, rather than as operators on collections. Likewise
for the binary logical operators.\\

\begin{tabular}{lll}
$<expression>$ & ::= & $<bracketed\_expression>$ $|$ $<equality\_expression>$ $|$\\
   &  & $<logical\_expression>$ $|$ $<factor\_expression>$ \\
$<bracketed\_expression>$ & ::= &  ``(" $<expression>$  ``)" \\
$<logical\_expression>$ & ::= & $<expression>$ $<logical\_op>$ $<expression>$  \\
$<equality\_expression>$ & ::= & $<factor\_expression>$ $<equality\_op>$ $<factor\_expression>$  \\
$<factor\_expression>$ & ::= & $<basic\_expression>$ $<factor\_op>$ $<factor\_expression>$ $|$\\
  & &  $<factor2\_expression>$  \\
$<factor2\_expression>$ & ::= & $<expression>$  ``-$>$any()" $|$\\
  & &  $<expression>$  ``-$>$size()" $|$\\
   &  &  $<expression>$ ``-$>$isDeleted()" $|$\\
  & & $<expression>$  ``-$>$exists(" $<identifier>$  ``$|$" $<expression>$  ``)" $|$\\
   &  &  $<expression>$  ``-$>$exists1(" $<identifier>$  ``$|$" $<expression>$  ``)" $|$\\
   &  &  $<expression>$  ``-$>$exists(" $<expression>$  ``)" $|$\\
   &  &  $<expression>$  ``-$>$exists1(" $<expression>$  ``)" $|$\\
   &  &  $<expression>$  ``-$>$forAll(" $<expression>$  ``)" $|$\\
   &  &  $<expression>$  ``-$>$select(" $<expression>$  ``)" $|$\\
   &  &  $<expression>$  ``-$>$reject(" $<expression>$  ``)" $|$ \\
   &  &  $<basic\_expression>$  \\
$<basic\_expression>$ & ::= & $<set\_expression>$ $|$ $<sequence\_expression>$ $|$ \\
  &  & $<call\_expression>$ $|$ $<array\_expression>$ $|$\\
  &  & $<identifier>$ $|$ $<value>$ \\
$<set\_expression>$ & ::= & ``$\{$" $<fe\_sequence>$ ``$\}$" \\ 
$<sequence\_expression>$ & ::= & ``$Sequence\{$" $<fe\_sequence>$ ``$\}$" \\ 
$<call\_expression>$ & ::= & $<identifier>$ ``(" $<fe\_sequence>$ ``)" \\ 
$<array\_expression>$ & ::= & $<identifier>$ ``[" $<fe\_sequence>$ ``]" 
\end{tabular}\\

A $logical\_op$ is one of {\tt =>}, {\tt \&}, {\tt or}.
An $equality\_op$ is one of $=$, $/=$, $>$, $<$, 
$<:$ (subset-or-equal), $<=$, $>=$, $:$, $/:$ (not-in).
A $factor\_op$ is one of $+$, $/$, $*$, $-$, $\backslash/$
(union), $\cat$ (concatenation of sequences), 
$/\backslash$ (intersection). An $fe\_sequence$ is a 
comma-separated sequence of factor expressions.
Identifiers can contain ``.".

\section{Activity syntax of UML-RSDS}
\label{actbnf}

The following concrete syntax is used for a subset of
UML structured activities:\\

\begin{tabular}{lll}
$<statement>$ & ::= & $<loop\_statement>$ $|$ $<creation\_statement>$ \\
   &  & $<conditional\_statement>$ $|$ $<sequence\_statement>$ $|$ \\
   &  & $<basic\_statement>$ \\
$<loop\_statement>$ & ::= & ``while" $<expression>$ ``do" $<statement>$ $|$  \\
    &  & ``for" $<expression>$ ``do" $<statement>$ \\
$<conditional\_statement>$ & ::= & ``if" $<expression>$ ``then" $<statement>$ \\
   &  & ``else" $<basic\_statement>$  \\
$<sequence\_statement>$ & ::= & $<statement>$ ``;" $<statement>$\\
$<creation\_statement>$ & ::= & $<identifier>$ ``:" $<identifier>$\\
$<basic\_statement>$ & ::= & $<basic\_expression>$ ``:=" $<expression>$ $|$ ``skip" $|$\\
  & & ``return" $<expression>$ $|$ ``(" $<statement>$  ``)" $|$ \\
  &  & $<call\_expression>$ \\
\end{tabular}


\small

\begin{thebibliography}{99}



 



\bibitem{modelmigrationcase} M. Herrmannsdoerfer,
{\em GMF: A Model Migration Case for the 
Transformation Tool Contest}, in \cite{ttc2011eptcs}, 
2011.





 

\bibitem{Lano08cdd}
K. Lano, 
{\em Constraint-Driven Development}, 
Information and Software Technology, 50,
2008, pp. 406--423. 

 




\bibitem{Lano11icmt} K. Lano, S. Kolahdouz-Rahimi,
{\em Model-Driven Development of Model
Transformations},
ICMT 2011.

\bibitem{Lano11icsea} K. Lano, S. Kolahdouz-Rahimi,
{\em Model Transformation Design Patterns},
ICSEA 2011.











\bibitem{ttc2011eptcs} Van Gorp, Pieter, Mazanek, Steffen,
and Rose, Louis,
{\em TTC 2011: Fifth Transformation Tool Contest,
Z\"urich, Switzerland, June 29-30 2011, Post-Proceedings},
EPTCS, 2011.

\end{thebibliography}
\end{document}